\documentclass[12pt]{article}

\usepackage{graphicx}
\usepackage{amsmath}

\newcommand{\ket}[1]{| #1 \rangle}
\newcommand{\bra}[1]{\langle #1 |}
\newcommand{\hcs}[1]{#1^\dagger #1}

\begin{document}
\begin{center}
{\Large\bf Nonunitary quantum circuit}
\vskip .6 cm
Hiroaki Terashima$^{1,2}$ and Masahito Ueda$^{1,2}$
\vskip .4 cm
{\it $^1$Department of Physics, Tokyo Institute of Technology,\\
Tokyo 152-8551, Japan} \\
{\it $^2$CREST, Japan Science and Technology Corporation (JST),\\
Saitama 332-0012, Japan}
\vskip .6 cm
\end{center}

\begin{abstract}
A quantum circuit is generalized to a nonunitary one
whose constituents are nonunitary gates
operated by quantum measurement.
It is shown that a specific type of
one-qubit nonunitary gates,
the controlled-\textsc{not} gate, as well as
all one-qubit unitary gates
constitute a universal set of gates
for the nonunitary quantum circuit,
without the necessity of introducing ancilla qubits.
A reversing measurement scheme is used to improve
the probability of successful nonunitary gate operation.
A quantum \textsc{nand} gate and Abrams-Lloyd's nonlinear
gate are analyzed as examples.
Our nonunitary circuit can be used to reduce
the qubit overhead needed
to ensure fault-tolerant quantum computation.
\end{abstract}

\begin{flushleft}
{\footnotesize
}
\end{flushleft}

\section{Introduction}
Quantum computation~\cite{NieChu00} is usually
described by unitary operations
because the time evolution of a closed system
is described by unitary transformations.
However, real systems interact with the environment,
which entails decoherence and errors in quantum computation.
To cope with the problem of decoherence,
quantum error-correcting schemes~\cite{Shor95,Steane96,LMPZBDSW96}
have been proposed in which redundant qubits are introduced
to ensure fault tolerance.
Unfortunately, this qubit overhead is too demanding,
since the number of available qubits
will be severely restricted in the foreseeable future.
To circumvent this problem,
a probabilistic quantum error-correcting scheme
without redundancy has recently been proposed~\cite{KoaUed99}
using a reversing measurement scheme~\cite{UeImNa96}.
This scheme involves quantum measurement
and is therefore described by nonunitary operations.

In this paper, we explore the possibility of
a general quantum information processing
based on nonunitary operations, i.e., quantum circuits
that involve not only unitary but also nonunitary gates,
the latter of which are implemented by quantum measurements.
In a sense, our nonunitary quantum circuit is
a generalization of the conventional unitary quantum circuit,
because the latter also invokes quantum measurement
at the end of computation.
However, in our scheme measurements are exploited
not only at the end but in the course of computation.
Of course, even in the usual quantum computer,
projective measurements are routinely used during computation.
For example,
Knill, Laflamme, and Milburn~\cite{KnLaMi01} have shown that
projective measurements can eliminate the need
for nonlinear couplings in an optical quantum computer.
Gilchrist et al.~\cite{GMMN03} have also shown that
an atomic measurement in an optical quantum computer
corresponds to a nonunitary operator that
is optically nonlinear and is approximately unitary
in a Hilbert subspace for a single mode.
Moreover,
Raussendorf and Briegel~\cite{RauBri01}, and
Nielsen~\cite{Nielse01} have recently proposed
two different schemes of quantum computation that
consist entirely of projective measurements.
Such measurements are intended to simulate unitary gates,
and thus unitary operators connect
the output states with the input states.
In contrast, nonunitary operators
are the connectors in our nonunitary quantum circuit,
based on a general framework of quantum measurement.
In other words,
\emph{our gates are nonunitary at the logical level
as well as at a physical level.}
This point emerges more clearly by comparing our circuit
with an optical quantum computer~\cite{RGMMG03}
that utilizes coherent states $\{\ket{-\alpha},\ket{\alpha}\}$
as the qubit states $\{\ket{0},\ket{1}\}$.
These qubit states are only approximately
orthogonal for large $|\alpha|$,
since $\langle 0|1 \rangle=\exp(-2|\alpha|^2)$.
Due to this non-orthogonality,
a unitary logical gate is physically implemented by
a nonunitary gate in a wider Hilbert space
using projective measurements.
For example, the Hadamard gate $H$,
which acts as
$H\ket{0}=\left(\ket{0}+\ket{1}\right)/\sqrt{2}$
and
$H\ket{1}=\left(\ket{0}-\ket{1}\right)/\sqrt{2}$,
should be nonunitary at the physical level
if the coherent-state qubit is used
because $\bra{0}\hcs{H}\ket{1}\neq\langle 0|1 \rangle$,
while the Hadamard gate itself is unitary at the logical level.

A natural question then arises as to whether or not
a universal set of nonunitary gates exists
for the nonunitary quantum circuit,
since it is well known that a set of unitary gates is universal
for the unitary quantum circuit~\cite{Deutsc89,DiVinc95,%
SlWeBa95,LDBE95,BBCDMS95,BMPRV99}.
We will show that a set of nonunitary gates
is universal for the nonunitary quantum circuit,
without the necessity of introducing ancilla qubits.
As a consequence of invoking quantum measurement,
the nonunitary gate operation is
necessarily probabilistic;
however, \emph{we can be sure
whether or not the gate operation is successful.}
We will discuss a reversing measurement scheme to increase
the probability of successful nonunitary gate operation
to the maximum allowable value.
However, the total probability of successful operation
of a nonunitary quantum circuit decreases exponentially
with the number of nonunitary gates.
This is a tradeoff for the reduction of qubit overhead,
because unsuccessful measurements destroy the quantum state
and halt the computation.
We will show that
if we could apply nonunitary gates with unit probability
by some quantum dynamics,
we could solve \textbf{NP}-complete problems in polynomial time.
This parallels the result
shown by Abrams and Lloyd~\cite{AbrLlo98}, in which
a hypothetical nonlinear quantum theory implies a
polynomial-time solution for \textbf{NP}-complete problems.

This paper is organized as follows.
Section~\ref{sec:nonuni} formulates nonunitary gates and
Sec.~\ref{sec:univ} discusses a universal set of nonunitary gates
for the nonunitary quantum circuit.
Section~\ref{sec:reverse} shows
a reversing measurement scheme to increase
the probability of success of a nonunitary gate.
Section~\ref{sec:example} considers
two examples of nonunitary gates:
a quantum \textsc{nand} gate and Abrams-Lloyd's nonlinear gate.
Section~\ref{sec:conclude} summarizes our results.

\section{\label{sec:nonuni}Nonunitary Gates}
We first define the nonunitary gate
as a generalization of the unitary gate.
A unitary gate is described as
$\ket{\psi} \to U \ket{\psi}$,
where $U$ is a unitary operator satisfying
$U^\dagger U=UU^\dagger=I$,
with $I$ being the identity operator.
In the computational basis for $n$ qubits,
the unitary operator $U$ is represented by
a complex-valued $2^n\times 2^n$ matrix
that satisfies the unitary condition~\cite{NieChu00}.
We define a nonunitary gate operation as
\begin{equation}
\ket{\psi} \to
\frac{N\ket{\psi}}{\sqrt{\bra{\psi}\hcs{N}\ket{\psi}}},
\label{nonunit}
\end{equation}
where $N$ is a nonunitary operator to be specified later.
In the computational basis for $n$ qubits,
$N$ is represented by a complex-valued $2^n\times 2^n$ matrix,
without being subject to the unitary condition.
Since a linear operation in a finite Hilbert space
is always bounded and
the normalization of $N$ does not affect
the state after the gate operation,
we normalize $N$ so that
the maximum eigenvalue of $\hcs{N}$ is unity:
\begin{equation}
\max_{\ket{\psi}} \,\bra{\psi}\hcs{N}\ket{\psi} =1.
\label{normal}
\end{equation}

To implement this nonunitary gate,
we utilize a general framework of quantum measurement,
in which a general measurement is described
by a set of measurement operators $\{M_m\}$~\cite{DavLew70}.
If the system is initially in a state $\ket{\psi}$,
the probability for outcome $m$ is given by
$p(m)=\bra{\psi}\hcs{M_m}\ket{\psi}$,
and the corresponding postmeasurement state is given by
\begin{equation}
\frac{M_m\ket{\psi}}{\sqrt{\bra{\psi}\hcs{M_m}\ket{\psi}}}.
\end{equation}
Since the total probability $\sum_m p(m)$ is $1$,
the measurement operators must satisfy
$\sum_m \hcs{M_m}=I$.
This means that
all the eigenvalues of
$\hcs{M_m}$ must be less than or equal to $1$.
That is, for any $\ket{\psi}$,
\begin{equation}
\bra{\psi}\hcs{M_m}\ket{\psi} \le 1.
\label{measure}
\end{equation}

Using this general measurement,
we implement the nonunitary gate $N$ as follows:
We perform a measurement $\{M_0, M_1\}$
with two outcomes, $0$ and $1$, such that
\begin{equation}
   M_0 =c\, N , \qquad M_1=\sqrt{I-\hcs{M_0}},
\label{meaop}
\end{equation}
where $c$ is a normalization constant.
It follows from Eqs.~(\ref{normal}) and (\ref{measure})
that $|c|\le 1$.
We assume that the successful measurement
corresponds to outcome $0$ and
the unsuccessful measurement corresponds to outcome $1$.
With the probability given by
\begin{equation}
p(\ket{\psi};c)=\bra{\psi}\hcs{M_0}\ket{\psi}
=|c|^2\bra{\psi}\hcs{N}\ket{\psi},
\label{succprob}
\end{equation}
the measurement is successful
and then the state of the system becomes
\begin{equation}
\ket{\psi} \to
\frac{M_0\ket{\psi}}{\sqrt{\bra{\psi}\hcs{M_0}\ket{\psi}}}=
\frac{N\ket{\psi}}{\sqrt{\bra{\psi}\hcs{N}\ket{\psi}}}.
\end{equation}
Comparing this equation with definition (\ref{nonunit}),
we find that this measurement implements
a nonunitary gate $N$ with
the probability of success $p(\ket{\psi};c)$
given in Eq.~(\ref{succprob}).
This probability is less than or equal to $|c|^2$
from Eq.~(\ref{normal}).
\emph{While the measurement may be unsuccessful,
we can be sure whether or not the gate operation
is successful by checking the measurement outcome.}
In terms of the quantum operations formalism~\cite{NieChu00},
the nonunitary gate $N$ is described by a quantum operation:
\begin{equation}
{\mathcal E}(\ket{\psi}\bra{\psi})=
M_0\ket{\psi}\bra{\psi}M_0^\dagger.
\end{equation}
Since this quantum operation does not include
a summation over the measurement outcomes,
a pure state remains pure
during the gate operation~\cite{Taraso02}.

While the constant $c$ does not affect
the postmeasurement state,
it does affect the probability of success.
The maximal probability of success is attained
by the measurement with $|c|=1$.
When this optimal measurement is not available,
we can still improve the probability of success
so that it is arbitrarily close to
the maximum allowable value
by applying a reversing measurement scheme
to a non-optimal measurement $|c|<1$,
as will be shown later.

There are, however, two problems with the nonunitary gate.
First, if $\det N$ is zero,
then there exists a state $\ket{\psi_W}$
such that $N\ket{\psi_W}=0$.
For this state, the measurement never succeeds
because $p(\ket{\psi_W};c)=0$.
We can circumvent this problem by excluding
the wrong states from the input state, or
by choosing $N$ such that $\det N$ is nonzero.
If $\det N\neq0$, the gate $N$ is said to be
logically reversible~\cite{UeImNa96}
in the sense that the input state can be
calculated from the output state.
In other words, a logically reversible gate preserves
all pieces of information about the input state
during the gate operation.
An example of the logically \emph{irreversible} gate is
the projective measurement, by the action of which
the information about the states orthogonal to
the projector is completely lost.
The second problem is that
the total probability of success of
a quantum circuit involving nonunitary gates
decays exponentially with the number of nonunitary gates,
since an unsuccessful measurement in an intermediate gate
forces us to restart from the first gate.
Nevertheless,
a nonunitary quantum circuit has the advantage of
reducing the number of qubits in some situations.
We shall discuss these two issues below,
using a quantum \textsc{nand} gate.

\section{\label{sec:univ}Universality}
We consider a universal set of gates
for the nonunitary quantum circuit both with and without
ancilla qubits.

\subsection{With ancilla qubits}
An arbitrary quantum measurement can be simulated
by the projective measurement and unitary operation
with the use of ancilla qubits.
Therefore, if ancilla qubits are available,
the controlled-\textsc{not} (\textsc{cnot}) gate,
all one-qubit unitary gates,
and the one-qubit projective measurement constitute
a universal set for the nonunitary quantum circuit.
We begin by proving this theorem.

\begin{figure}
\begin{center}
\includegraphics[scale=0.6]{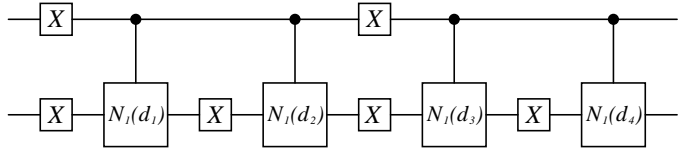}
\end{center}
\caption{\label{fig1}Circuit for a $D(d_1,d_2,d_3,d_4)$ gate.}
\end{figure}

\begin{figure}
\begin{center}
\includegraphics[scale=0.6]{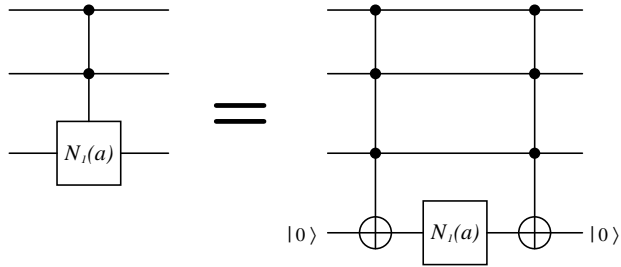}
\end{center}
\caption{\label{fig2}Circuit for a $C^2[N_1(a)]$ gate
using an ancilla qubit (represented by the bottom line).}
\end{figure}

\begin{figure}
\begin{center}
\includegraphics[scale=0.6]{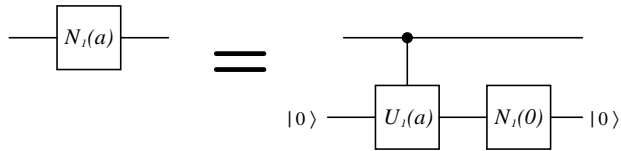}
\end{center}
\caption{\label{fig3}Circuit for a $N_1(a)$ gate
using an ancilla qubit (represented by the bottom line).}
\end{figure}

Consider a $2^n\times 2^n$ nonunitary matrix $N$
representing a nonunitary quantum circuit for $n$ qubits.
We then make the singular value decomposition of $N$,
\begin{equation}
N=U\,D(d_1,d_2,\cdots,d_{2^n})\,V,
\end{equation}
where $U$ and $V$ are unitary matrices
and $D(d_1,d_2,\cdots,d_{2^n})$ is a diagonal matrix
whose diagonal components $\{d_i\}$ satisfy $0\le d_i\le 1$.
The unitary matrices $U$ and $V$ can be further
decomposed into \textsc{cnot} gates and one-qubit unitary gates,
since this set of gates is universal
for the unitary quantum circuit~\cite{BBCDMS95}.
We thus concentrate on the diagonal matrix
$D(d_1,d_2,\cdots,d_{2^n})$.
Using the \textsc{not} gate
$X=\left(\begin{smallmatrix} 0 & 1 \\
1 & 0 \end{smallmatrix}\right)$,
this matrix can be factorized into
$2^n$ matrices that have the form of $D(1,1,\cdots,1,d_i)$.
Each of these matrices corresponds to
a controlled-$N_1(a)$ gate with $n-1$ control qubits
(denoted by $C^{n-1}[N_1(a)]$),
where $N_1(a)$ is a one-qubit nonunitary gate given by
\begin{equation}
N_1(a)\equiv\left(\begin{array}{cc}
    1 & 0 \\
    0 & a
    \end{array}\right), \qquad0\le a<1.
\end{equation}
The case of $n=2$ is illustrated in Fig.~\ref{fig1}.
This $C^{n-1}[N_1(a)]$ gate can be implemented by
one $N_1(a)$ gate and two $C^n[X]$ gates (i.e.,
two \textsc{cnot} gates with $n$ control qubits)
with the help of an ancilla qubit prepared in state $\ket{0}$.
Figure~\ref{fig2} shows the case of $n=3$.
Since the $C^n[X]$ gate is unitary,
it can be decomposed into
\textsc{cnot} gates and one-qubit unitary gates~\cite{BBCDMS95}.
On the other hand, as shown in Fig.~\ref{fig3},
the $N_1(a)$ gate can be decomposed into
one $N_1(0)$ gate and one controlled-$U_1(a)$ gate
by using an ancilla qubit,
where $U_1(a)$ is a one-qubit unitary gate defined by
\begin{equation}
U_1(a)\equiv\left(\begin{array}{cc}
    a & \sqrt{1-a^2} \\
    \sqrt{1-a^2} & -a
    \end{array}\right).
\end{equation}
Since the $N_1(0)$ gate corresponds to
the one-qubit projective measurement $\ket{0}\bra{0}$,
the theorem is proved.

\subsection{Without ancilla qubits}
In view of necessity of reducing the number of qubits,
we next consider the case
where no ancilla qubits are available.
In this case, we cannot find a set of gates with which
to exactly construct an arbitrary nonunitary circuit.
We therefore apply a definition of universality
in a broad sense
in which two gates are regarded as identical
if they differ only by a normalization factor.
Note that the normalization of a nonunitary
gate does not affect the state after the gate operation,
though it affects the probability of success.
With this proviso,
we here prove that if ancilla qubits are not available,
the \textsc{cnot} gate,
all one-qubit unitary gates, and the $N_1(a)$ gates ($0\le a<1$)
constitute a universal set for the nonunitary circuit.

\begin{figure}
\begin{center}
\includegraphics[scale=0.6]{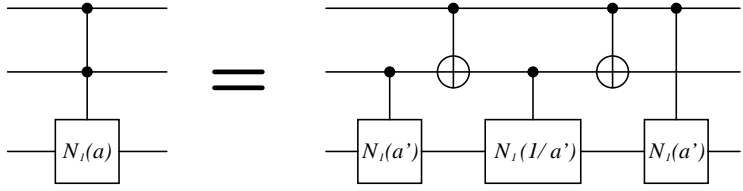}
\end{center}
\caption{\label{fig4}Circuit for a $C^2[N_1(a)]$ gate
with no ancilla qubit, where $a'=\sqrt{a}$.}
\end{figure}

\begin{figure}
\begin{center}
\includegraphics[scale=0.6]{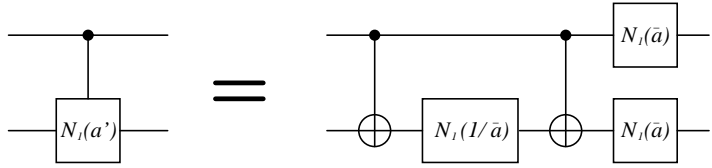}
\end{center}
\caption{\label{fig5}Circuit for a
controlled-$N_1(a')$ gate with no ancilla qubit,
where $\bar{a}=\sqrt{a'}$.}
\end{figure}

The proof goes as follows.
As shown in the preceding section,
an arbitrary  nonunitary matrix can be decomposed into
the $C^{n-1}[N_1(a)]$ gates and unitary matrices.
When $a\neq0$,
each $C^{n-1}[N_1(a)]$ gate can be further decomposed
into controlled-$N_1(a')$ gates,
controlled-$N_1(1/a')$ gates, and \textsc{cnot} gates
without ancilla qubits as in the unitary case~\cite{BBCDMS95},
where $a'=a^{\frac{1}{2^{n-2}}}$.
Figure~\ref{fig4} illustrates the case of $n=3$.
Moreover, as shown in Fig.~\ref{fig5},
the controlled-$N_1(a')$ gate can be implemented
by two $N_1(\bar{a})$ gates,
one $N_1(1/\bar{a})$ gate, and two \textsc{cnot} gates,
where $\bar{a}=\sqrt{a'}$.
Similarly, the controlled-$N_1(1/a')$ gate can
be implemented by two $N_1(1/\bar{a})$ gates,
one $N_1(\bar{a})$ gate, and two \textsc{cnot} gates.
Therefore, we are left with the $N_1(\bar{a})$ and $N_1(1/\bar{a})$ gates
($0< \bar{a}<1$), apart from the \textsc{cnot} gate and
the wire $N_1(1)$.
However, the $N_1(1/\bar{a})$ gate is not a gate
in the strict sense, because it does not satisfy
the normalization condition (\ref{normal}) due to $1/\bar{a}>1$.
We must renormalize it using the identification
up to a normalization factor.
Namely, via the equation
\begin{equation}
 N_1(1/a) \propto X\,N_1(a)\,X,
\label{invna}
\end{equation}
we express the $N_1(1/\bar{a})$ gate by
one $N_1(\bar{a})$ gate and two \textsc{not} gates.

On the other hand, when $a=0$, the $C^{n-1}[N_1(a)]$ gate
becomes the $D(1,1,\cdots,1,0)$ gate,
which is paired with the $D(0,0,\cdots,0,1)$ gate
through the relation $\hcs{M_0}+\hcs{M_1}=I$.
This means that the unsuccessful operation of
the $D(0,0,\cdots,0,1)$ gate is identical to
the successful operation of the $C^{n-1}[N_1(0)]$ gate,
and vice versa.
We thus construct the $D(0,0,\cdots,0,1)$ gate
instead of the $C^{n-1}[N_1(0)]$ gate.
Note that the $D(0,0,\cdots,0,1)$ gate corresponds to
the $n$-qubit projective measurement $\ket{11\cdots1}\bra{11\cdots1}$.
Since the one-qubit projective measurement $\ket{1}\bra{1}$
corresponds to the $X N_1(0) X$ gate,
the operation of the $D(0,0,\cdots,0,1)$ gate can be implemented by
the action of the $X N_1(0) X$ gate on each qubit.
Consequently, we can construct any nonunitary quantum circuit
from the \textsc{cnot} gate, all one-qubit unitary gates,
and the $N_1(a)$ gates with $0\le a<1$.

We finally show that any $N_1(a)$ gate can be \emph{approximated}
to arbitrary accuracy by only two fixed nonunitary gates
together with the \textsc{not} gate.
For a real number $\alpha$ and an irrational number $\gamma$,
we consider the $N_1(\alpha)$ and $N_1(\alpha^\gamma)$ gates.
Note that for any real numbers $a$ and $\epsilon$,
there exist integers $m$ and $l$ such that
\begin{equation}
  \Bigl|\,\log_\alpha a
 -(m\gamma+l)\,\Bigr|<\epsilon.
\end{equation}
Using these $m$ and $l$, the $N_1(a)$ gate
is approximately written as
\begin{equation}
 N_1(a) \sim \left[ N_1(\alpha^\gamma) \right]^m
\left[ N_1(\alpha) \right]^l.
\end{equation}
If $m<0$ or $l<0$, we use Eq.~(\ref{invna})
to make the power positive;
ignoring the normalization factor, we obtain
\begin{eqnarray}
\left[ N_1(\alpha) \right]^l &\propto&
X \left[ N_1(\alpha) \right]^{-l}X, \\
\left[ N_1(\alpha^\gamma) \right]^m &\propto&
X \left[ N_1(\alpha^\gamma) \right]^{-m}X.
\end{eqnarray}
In this way, we can approximate any $N_1(a)$ gate by only
the $N_1(\alpha)$ and $N_1(\alpha^\gamma)$ gates
together with the \textsc{not} gate.

\section{\label{sec:reverse}Optimization by Reversing Measurement}
\begin{figure}
\begin{center}
\includegraphics[scale=0.55]{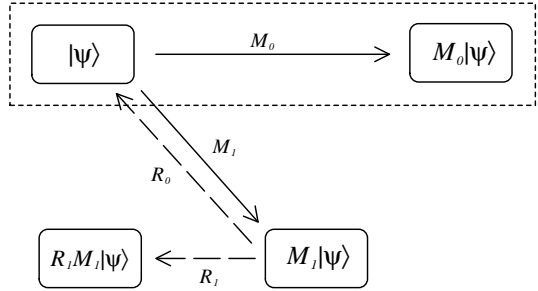}
\end{center}
\caption{\label{fig6}A reversing measurement scheme.
If the measurement $\{M_0,M_1\}$ (solid arrows) fails,
then a reversing measurement $\{R_0,R_1\}$ (dashed arrows)
probabilistically reverts the postmeasurement state
$M_1\ket{\psi}$ back to the original state $\ket{\psi}$.}
\end{figure}

To implement a nonunitary gate $N$,
we must prepare the measurement $\{M_0,M_1\}$
defined in Eq.~(\ref{meaop}).
The probability of success of this measurement
depends on the value of $c$.
When the optimal measurement $|c|=1$ is not available,
the probability of success of the gate operation is reduced.
However, we can improve the probability of success so that
it is arbitrarily close to the maximum allowable value
by applying a reversing measurement scheme~\cite{UeImNa96}
to a non-optimal measurement $|c|<1$ (see Fig.~\ref{fig6}).
More specifically,
if the measurement $\{M_0,M_1\}$ fails,
we perform another measurement
$\{R_0,R_1\}$ that satisfies
\begin{equation}
 R_0 \,M_1= q I, \qquad
R_1=\sqrt{I-\hcs{R_0}},
\label{revmea}
\end{equation}
where $q$ is a constant.
Note that $R_0$ exists when $|c|<1$
and is proportional to $M_1^{-1}$.
Therefore,
if this measurement is successful,
the postmeasurement state becomes
the original one $\ket{\psi}$ and
we can then try the measurement $\{M_0,M_1\}$ again
to increase the probability of success of
the nonunitary gate operation.
Of course, the reversing measurement
$\{R_0,R_1\}$ also fails with a nonzero probability.
The joint probability for $M_1$ followed by
$R_0$ is given by $|q|^2$, which does not
depend on the measured state $\ket{\psi}$.
Note that $|q|$ cannot be set to $1$,
since the maximum eigenvalue of $\hcs{R_0}$
is $|q|^2/(1-|c|^2)$ which must be less than $1$.
In order for $R_0$ to be
a measurement operator,
Eq.~(\ref{measure}) requires that
\begin{equation}
0<|q|\le\sqrt{1-|c|^2}.
\end{equation}

Using the reversing measurement once,
the probability of success of the nonunitary gate operation
increases to $p(\ket{\psi};c)+|q|^2p(\ket{\psi};c)$,
where the first and second terms result from the process $M_0$ and
the process $M_1\to R_0\to M_0$, respectively.
The other processes, $M_1\to R_1$ and $M_1\to R_0\to M_1$, are
unsuccessful gate operations.
However, we can repeatedly perform the reversing measurement
when the process ends with $M_1$
in order to further increase the probability of success.
By repeating the reversing measurement, at most $k$ times
as long as the reversing measurement succeeds,
we increase the probability of success to
\begin{eqnarray}
\tilde{p}_{k}(\ket{\psi};c)
 &=& \left[1+|q|^2+\cdots+|q|^{2k} \right]\,p(\ket{\psi};c) \nonumber \\
 &=& \frac{1-|q|^{2k+2}}{1-|q|^2}\,p(\ket{\psi};c),
\end{eqnarray}
since we can repeat the process $M_1\to R_0$ $l$ times
with probability $|q|^{2l}$.
Substituting the maximum value $\sqrt{1-|c|^2}$ for $|q|$,
we find
\begin{equation}
 \tilde{p}_{k}(\ket{\psi};c) =
\frac{1-\left(1-|c|^2\right)^{k+1}}{|c|^2}\,
p(\ket{\psi};c).
\label{revpro}
\end{equation}
In the limit of $k\to\infty$, we obtain
\begin{equation}
\tilde{p}_{\infty}(\ket{\psi};c)=
\frac{1}{|c|^2}\,p(\ket{\psi};c)=p(\ket{\psi};1).
\end{equation}
This shows that even if the measurement to implement
a nonunitary gate is not optimal, i.e., $|c|<1$,
we can, in principle, increase the probability of success
so that it is arbitrarily close to
the optimal value with $|c|=1$
by utilizing the reversing measurement scheme,
provided that the optimal reversing measurement with
$|q|=\sqrt{1-|c|^2}$ is available.
Note that the optimal value with $|c|=1$ does not mean
a deterministic nonunitary gate,
as can be seen from Eq.~(\ref{succprob}).
We cannot increase the probability of success to $1$,
since the reversing measurement is
only successful in a probabilistic way.
Nevertheless, it is worthwhile to note that
the reversing measurement scheme can improve
the probability of success to some extent
if the original one is not optimal.

\section{\label{sec:example}Examples}
Finally we discuss two examples of the nonunitary gate.
The first one is a quantum \textsc{nand} gate,
which can reduce the work space required to
perform a specific type of quantum computation.
The second one is Abrams-Lloyd's nonlinear gate,
which could solve the \textbf{NP}-complete problem
in polynomial time
if its probability of success were $1$.

\subsection{Quantum \textmd{\textsc{nand}} gate}
\begin{table}
\begin{center}
\begin{tabular}{|c|c|} \hline
Inputs & Output \\ \hline
 0 \quad 0 & 1 \\
 0 \quad 1 & 1 \\
 1 \quad 0 & 1 \\
 1 \quad 1 & 0 \\ \hline
\end{tabular}
\end{center}
\caption{\label{tbl1}The truth table of the \textsc{nand} gate.}
\end{table}
The classical \textsc{nand} gate (Table~\ref{tbl1})
is a universal gate for irreversible classical computation.
However, a quantum version of this gate cannot be a unitary gate,
due to the irreversibility of the \textsc{nand} operation.
In contrast,
we can make a quantum version of the \textsc{nand} gate
as a nonunitary gate.

Consider a two-qubit nonunitary gate
represented in a computational basis,
\[
\ket{00}=
\left(\begin{array}{c} 1 \\ 0 \\ 0 \\ 0 \end{array}\right), \quad
\ket{01}=
\left(\begin{array}{c} 0 \\ 1 \\ 0 \\ 0 \end{array}\right), \quad
\ket{10}=
\left(\begin{array}{c} 0 \\ 0 \\ 1 \\ 0 \end{array}\right), \quad
\ket{11}=
\left(\begin{array}{c} 0 \\ 0 \\ 0 \\ 1 \end{array}\right),
\]
as
\begin{equation}
N=\frac{1}{\sqrt{3}}\left(\begin{array}{cccc}
    0  & 0 & 0 & 1 \\
    0  & 0 & 0 & 0 \\
    1  & 1 & 1 & 0 \\
    0  & 0 & 0 & 0
  \end{array}\right).
\end{equation}
This gate transforms the computational basis as
$N \ket{00}=N\ket{01}=N\ket{10}=\ket{10}$
and $N \ket{11}=\ket{00}$, which
yields the truth table of the \textsc{nand} gate
as in Table~\ref{tbl1}
if the second qubit of the output state is ignored.
We thus call $N$ a quantum \textsc{nand} gate.
Note that the second qubit always
becomes $\ket{0}$ after the gate operation
in order not to become entangled with the first qubit.
To implement this nonunitary gate,
we prepare a measurement $\{M_0, M_1\}$
with two outcomes, $0$ and $1$,
as in Eq.~(\ref{meaop}).
For the states in the computational basis,
the probabilities of success are equal,
$p(\ket{x};c)=|c|^2/3$ for $x\in\{00,01,10,11\}$.
When the initial state is
\begin{equation}
\ket{\psi_{\max}}=
\frac{\ket{00}+\ket{01}+\ket{10}}{\sqrt{3}},
\end{equation}
the probability of success becomes maximal
due to constructive interference:
$p(\ket{\psi_{\max}};c)=|c|^2$.
Therefore, $c$ must satisfy $0< |c| \le 1$.
Otherwise $\hcs{M_1}=I-\hcs{M_0}$
fails to be positive semidefinite.
An explicit form of $M_1$ is given by
\begin{equation}
 M_1 =\frac{1}{3}\left(\begin{array}{cccc}
    2+a  & -1+a & -1+a & 0  \\
   -1+a  &  2+a & -1+a & 0  \\
   -1+a  & -1+a &  2+a & 0  \\
      0  &    0 &    0 & 3b
  \end{array}\right),
\end{equation}
where $a=\sqrt{1-|c|^2}$ and $b=\sqrt{1-(|c|^2/3)}$.

On the other hand,
the minimum probability of success is
\begin{equation}
\min_{\ket{\psi}} \,p(\ket{\psi};c)=0,
\end{equation}
since the minimum eigenvalue of $\hcs{N}$ is zero.
Two eigenvectors correspond to the zero eigenvalue,
\begin{equation}
 \frac{\ket{00}-\ket{01}}{\sqrt{2}}, \qquad
 \frac{\ket{01}-\ket{10}}{\sqrt{2}}.
\end{equation}
This means that the measurement never succeeds for the states
in the two-dimensional subspace spanned by these vectors,
due to destructive interference.
For example, $N\left(\ket{00}-\ket{01}\right)/\sqrt{2}=0$.
Note that $N$ is not logically reversible because $\det N=0$.
When using $N$, we must exclude these wrong states
from the input state.

Since $M_1$ is logically reversible ($\det M_1\neq0$)
in the non-optimal case $|c|<1$,
the reversing measurement scheme can be utilized
to improve the probability of success.
We can thus perform the reversing measurement
$\{R_0, R_1\}$ defined by Eq.~(\ref{revmea}).
The explicit form of $R_0$ is
\begin{equation}
R_0=\frac{q}{3a}\left(\begin{array}{cccc}
   1+2a  &  1-a &  1-a & 0 \\
    1-a  & 1+2a &  1-a & 0 \\
    1-a  &  1-a & 1+2a & 0 \\
      0  &    0 &    0 & 3a/b
  \end{array}\right).
\end{equation}
It is easy to confirm that $R_0M_1$ is equal to $qI$
and that the eigenvalues of $\hcs{R_0}$ are
less than or equal to $1$
if $|q|\le\sqrt{1-|c|^2}$.
The reversing measurement scheme then increases
the probability of success as in Eq.~(\ref{revpro})
if $|q|=\sqrt{1-|c|^2}$.
For the states in the computational basis
$\ket{x}$ with $x\in\{00,01,10,11\}$,
we obtain
\begin{equation}
\tilde{p}_{k}(\ket{x};c)
=\frac{1-\left(1-|c|^2\right)^{k+1}}{3},
\end{equation}
and for the maximally successful state we obtain
\begin{equation}
\tilde{p}_{k}(\ket{\psi_{\max}};c)
=1-\left(1-|c|^2\right)^{k+1}.
\end{equation}
In the limit of $k\to\infty$,
they become the maximum allowable values,
$1/3$ and $1$, respectively.

As an application of the quantum \textsc{nand} gate,
we consider computing
\begin{equation}
 \frac{1}{\sqrt{2^n}}\sum_{x=0}^{2^n-1} \ket{x}\ket{0}
\to \frac{1}{\sqrt{2^n}}\sum_{x=0}^{2^n-1}\ket{x}\ket{f(x)}
\label{problem}
\end{equation}
for a given function $f(x)$.
(Consider, for example, the modular exponentiation
in Shor's algorithm~\cite{Shor97}.)
In conventional quantum computers,
we build up a unitary quantum circuit for
this computation by the following steps~\cite{NieChu00}:
({\romannumeral 1}) We construct an irreversible classical circuit
to calculate $x\to f(x)$ using classical \textsc{nand} gates,
since these gates are universal in classical computation.
({\romannumeral 2}) We replace the classical \textsc{nand} gates
with classical Toffoli gates
to make this classical circuit reversible,
by adding ancilla bits.
({\romannumeral 3}) We translate this reversible classical circuit
into a quantum one by replacing
the classical Toffoli gates with quantum Toffoli gates.
Note that the resultant circuit needs more qubits than
the irreversible classical circuit,
due to step ({\romannumeral 2}).

Because all coefficients of
the linear combination in Eq.~(\ref{problem}) are positive,
no destructive interference occurs
in operating the quantum \textsc{nand} gate.
We thus utilize the quantum \textsc{nand} gate
to reduce the number of qubits needed.
Instead of steps ({\romannumeral 2}) and ({\romannumeral 3}),
we directly replace the classical \textsc{nand} gates with
quantum \textsc{nand} gates.
This procedure allows us to reduce the number of qubits
needed to perform calculation (\ref{problem}),
because the quantum \textsc{nand} gate is a two-qubit gate,
unlike the quantum Toffoli gate.
However, the quantum \textsc{nand} gate is probabilistic,
since it is implemented by quantum measurement.
When all the classical \textsc{nand} gates are replaced
with quantum \textsc{nand} gates,
the probability of success becomes exponentially small
as the number of \textsc{nand} gates increases.
Thus, in practice, we replace only
some classical \textsc{nand} gates with quantum ones.
After dividing the function $f$ into
two functions, $g_1$ and $g_2$, i.e.,
\begin{equation}
x\longrightarrow g_1(x) \longrightarrow g_2(g_1(x))=f(x),
\end{equation}
we calculate $g_1$ using quantum \textsc{nand} gates
and $g_2$ using quantum Toffoli gates.
If $g_1$ contains $m$ quantum \textsc{nand} gates,
this method can save $m$ qubits
with the probability of success $(|c|^{2}/3)^m$.
By checking the measurement outcome,
we can be sure whether or not
the gate operations are successful.

\subsection{Abrams-Lloyd's gate}
Abrams and Lloyd~\cite{AbrLlo98} showed that
\textbf{NP}-complete problems could be solved
in polynomial time if quantum theory were nonlinear
at some level.
Although the nonlinearity of quantum theory
is hypothetical,
their work establishes a new link between a physical law
and the power of computing machines.
We here describe their nonlinear gate
as a nonunitary gate.

Let $F(x)$ be a function that maps an $n$-bit input
to a single bit $\{0,1\}$.
Given an oracle to calculate $F(x)$,
can we determine whether or not
there exists an input value $x$ for which $F(x)=1$?
The integer $s$ is defined as the number of such input values
and, to simplify the problem,
is assumed to be either $0$ or $1$.
In order to solve this \textbf{NP}-complete problem,
Abrams and Lloyd first prepare $n$ qubits $\ket{x}$ and
one flag qubit $\ket{F(x)}$ in an entangled state
\begin{equation}
 \ket{\psi_i}=\frac{1}{\sqrt{2^n}}\sum_{x=0}^{2^n-1}
  \ket{x}\ket{F(x)},
\end{equation}
using the usual quantum computer.
They then let a two-qubit quantum gate
perform a nonlinear transformation
\begin{eqnarray}
\frac{1}{\sqrt{2}}\left(\ket{00}+\ket{11}\right)
  &\longrightarrow &
 \frac{1}{\sqrt{2}}\left(\ket{01}+\ket{11}\right), \nonumber \\
\frac{1}{\sqrt{2}}\left(\ket{01}+\ket{10}\right)
  &\longrightarrow &
 \frac{1}{\sqrt{2}}\left(\ket{01}+\ket{11}\right), \label{nonlinear} \\
\frac{1}{\sqrt{2}}\left(\ket{00}+\ket{10}\right)
  &\longrightarrow &
 \frac{1}{\sqrt{2}}\left(\ket{00}+\ket{10}\right), \nonumber
\end{eqnarray}
successively on the first and flag qubits,
on the second and flag qubits,
and so on through the $n$th and flag qubits.
The final state is given by
\begin{equation}
\ket{\psi_f}=
\left(\frac{1}{\sqrt{2^n}}\sum_{x=0}^{2^n-1} \ket{x}\right)\ket{s},
\end{equation}
in which the flag qubit is not entangled with the first $n$ qubits.
Therefore, by measuring the flag qubit,
the answer $s$ is found in polynomial time.

It is easy to see that our nonunitary gate
can simulate Abrams-Lloyd's nonlinear gate as
\begin{equation}
 N_{\mathrm{AL}}=\frac{1}{\sqrt{6}}
  \left(\begin{array}{cccc}
   0  & -1 &  1 &  0 \\
   0  &  1 &  0 &  1 \\
   0  & -1 &  1 &  0 \\
   0  &  1 &  0 &  1
  \end{array}\right),
\end{equation}
even though the nonunitary gate is linear except for
the normalization factor.
This means that \textbf{NP}-complete problems
could be solved in polynomial time if
the $N_{\mathrm{AL}}$ gate could be applied with
probability $1$ by some quantum dynamics.
We can thus establish yet another new link between
a physical law and the power of computing machines.

Unfortunately, as discussed in the preceding sections,
the implementation of a nonunitary gate by quantum measurement
is intrinsically probabilistic.
Even if the implementing measurement is optimal ($|c|=1$),
the $N_{\mathrm{AL}}$ gate succeeds only
with probability $1/6$ for the states (\ref{nonlinear});
otherwise an unsuccessful operator, e.g.,
\begin{equation}
 M_1=\frac{1}{\sqrt{6}}
  \left(\begin{array}{cccc}
   \sqrt{6}  & 0 &  0 &  0 \\
   0  &  0 &  0 &  0 \\
   0  &  1 &  2 &  0 \\
   0  & -1 &  0 &  2
  \end{array}\right),
\end{equation}
is applied to the states.
When the number of qubits is $n$,
the total probability of success
decays exponentially as $(1/6)^n$.
To obtain a definite result,
we must repeat the algorithm $6^n$ times,
which consumes an exponential computation time.

\section{\label{sec:conclude}Conclusions}
We have formulated a nonunitary quantum circuit
having nonunitary gates operated by quantum measurement.
In contrast with recently proposed schemes on
quantum computation using
measurements~\cite{KnLaMi01,GMMN03,RauBri01,Nielse01,RGMMG03},
our gates utilize the nonunitarity fully
in the sense that
not only the physical implementation
but also the logical operation is nonunitary.
We have shown that the \textsc{cnot} gate,
a complete set of one-qubit unitary gates,
and the $N_1(a)$ gates constitute a universal set of gates for
the nonunitary quantum circuit
without the necessity of introducing ancilla qubits,
and have shown that a nonunitary gate can be optimized
by a reversing measurement scheme.
These results will be useful for the construction of
a quantum computer equipped with probabilistic error correction
by the reversing measurement.
More generally, the nonunitary quantum circuit can reduce
the number of qubits required to perform
some kinds of quantum computation,
as illustrated by the quantum \textsc{nand} gate.
Although we cannot reduce the number of qubits excessively,
due to the probabilistic nature of the nonunitary gate,
this approach would be useful for constructing a quantum computer
as long as the number of available qubits is severely restricted.
Moreover, apart from this practical interest,
there may be an academic interest in extending
quantum computation itself to include nonunitary operations.
At least, using Abrams-Lloyd's gate,
the nonunitary quantum computer can solve
\textbf{NP}-complete problems in polynomial time
(if the probabilistic nature is ignored),
whereas it is widely believed that
the usual unitary quantum computer cannot do so.
It would be interesting to quantify how much nonunitarity
is required to solve \textbf{NP}-complete problems
in polynomial time.

\section*{Acknowledgments}
This research was supported by a Grant-in-Aid
for Scientific Research (Grant No.~15340129) by
the Ministry of Education, Culture, Sports,
Science and Technology of Japan.


\end{document}